# Magnetic field-dependent dielectric constant in La$_{2/3}$Ca$_{1/3}$MnO$_3$

J. Rivas, J. Mira, B. Rivas-Murias, and A. Fondado
Departamento de Física Aplicada, Universidade de Santiago de Compostela
E-15782 Santiago de Compostela, Spain

J. Dec and W. Kleemann
Angewandte Physik, Universität Duisburg-Essen
D-47048 Duisburg, Germany

M. A. Señarís-Rodríguez
Departamento de Química Fundamental, Universidade da Coruña
E-15071 A Coruña, Spain

We report a rather high dependence of the dielectric permittivity on the magnetic field in La$_{2/3}$Ca$_{1/3}$MnO$_3$. The variation is maximum at around 270 K, little above the Curie temperature, $T_C$, and it reaches a 30% under only 0.5 T. We attribute this phenomenon to the space-charge or interfacial polarization produced between the insulator and the metallic regions segregated intrinsically in the material above $T_C$.

The interest in the control of the dielectric properties of a material by a magnetic field has given rise to an active research line in recent times.[1-3] All the findings have involved the participation of ferroelectricity in coexistence with magnetic ordering, coexistence that in the used materials takes place at relatively low temperatures.

Up to date, the magnetic control of the real part of the complex relative dielectric permittivity ($\varepsilon_r = \varepsilon'_r - i\varepsilon''_r$) -dielectric constant, $\varepsilon'_r$- has been achieved basically by two strategies: The first consists in finding compounds with strong coupling between ferroelectricity and magnetism (magnetoelectric effect). In this line, Kimura *et al.*[1] have found in TbMnO$_3$ single crystals variations of $\varepsilon'_r$ that go from 4% to 10%, depending on the selected crystal axis, under magnetic fields up to 9 T and at temperatures below 20 K. Hur *et al.*[2] have also found noticeable magnetodielectric effects in DyMn$_2$O$_5$, under magnetic fields up to 8 T at temperatures below 20 K. In both cases, the cause seems to be the switching of electric polarization in frustrated spin systems or the coupling of a magnetic transition with a dielectric transition. Very recently, Hemberger *et al.*[3] have reported an almost 500% variation in the spinel compound CdCr$_2$S$_4$, that shows a coupling between its ferromagnetic order and its relaxor ferroelectricity. As for the second strategy, it consists of the fabrication of composites or solid solutions of ferroelectric perovskites with magnetic order, like that created by Mitoseriu *et al.*[4] with PbFe$_{2/3}$W$_{1/3}$O$_3$ and PbTiO$_3$.

In this work we are reporting the finding of a dielectric permittivity notably dependent on the external magnetic field in the magnetic and non-ferroelectric compound La$_{2/3}$Ca$_{1/3}$MnO$_3$, a symbolic one in condensed matter physics, as it has starred miriads of scientific papers due to its appealing properties, like colossal magnetoresistance[5] and a plethora of other phenomena.[6] This finding cannot be classified into any of the aforementioned strategies. Moreover, the phenomenon is observed at room temperature and under relatively small magnetic fields.

The sample was prepared by a conventional solid-state reaction starting from stoichiometric amounts of $La_2O_3$, CaO, MnO and $MnO_2$, which were thoroughly mixed and grinded together, pressed at 5 Ton/cm$^2$ into pellets and finally fired at 1573 K for 48 hours with intermediate grindings. The sample was then cooled to room temperature at the rate of 1 K/min. The complex dielectric permittivity was measured with a parallel-plate capacitor coupled to a precision LCR meter Agilent 4284 A, capable to measure in frequencies ranging from 20 to $10^6$ Hz. The capacitor was mounted in an aluminium box refrigerated with liquid nitrogen, and incorporating a mechanism to control the temperature and the external magnetic field up to 0.5 T. The sample was prepared to fit in the capacitor, and gold was sputtered on their surfaces to ensure good electrical contact with the electrodes of the capacitor. The system was tested using a commercial $SrTiO_3$ sample, and it gave values similar to those reported in the literature.[7] Dielectric data were also taken with a dielectric spectrometer Solartron 1260 with interface 1296 in two ways: with a silver paste electrode and with the sample electroded with flash evaporation of Cu and subsequent sputtering of Au. The dc resistivity was measured by a standard four-probe technique using a homemade device.

Fig. 1 shows the temperature dependence of the real part of the relative dielectric permittivity, $\varepsilon'_r$, at several frequencies. It reaches values of ~$10^4$, evidencing the role of extrinsic effects.[8-10] The shape of the curves, with a well defined maximum at $T_m$ ~ 260 K, and the frequency dependence, indicate dielectric relaxation. It is, nevertheless, different from that observed in relaxor ferroelectrics[11] in two aspects, in principle: firstly, $T_m$ does not depend on frequency (this temperature is close to $T_C$, below which the material becomes ferromagnetically ordered after a first-order phase transition[12]). Secondly, standard dielectric relaxation involves strong frequency dependence below $T_m$, and smaller dependence above it (which is no more than a consequence of the thermal activated mobility of the dipolar species); the contrary is observed here. The application of an external and relatively small magnetic field is sufficient to cause strong variations in the dielectric response above $T_C$ (Fig. 2): it provokes an increase of about 35% under a magnetic field of 0.5 T around 280-300 K (Fig. 2, inset). For higher temperatures the percentage variation decreases and at around 320 K the magnetic dependence of $\varepsilon'_r$ is already small. On the other hand, the field dependence is almost negligible below $T_m$. Also, it is noteworthy the shift of $T_m$ to higher temperatures with increasing magnetic field.

The interpretation of this dielectric response must take into account the peculiar phase-separated electronic state of $La_{2/3}Ca_{1/3}MnO_3$ that arises above $T_C$.[13] Such a picture has served to explain satisfactorily the rich phenomenology observed in this compound. This phase-separated state, consisting basically of a mixture of insulating and metallic ferromagnetic regions, would cause space-charge or interfacial polarization, that would yield a dielectric response analogous to that generated by conducting regions in an insulating matrix.[7] The basic principle is the same that inspires the fabrication of some "artificial" dielectric systems, like cermets.[14] The qualitative difference is that in $La_{2/3}Ca_{1/3}MnO_3$ this mixture arises spontaneously in a natural way and it is *intrinsic* to the material, whereas in the systems known up to date the interfacial polarization has to be extrinsically produced by fabricating composites or mixtures of diverse compounds. In the case of $La_{2/3}Ca_{1/3}MnO_3$ the presence of these electronically segregated regions above the first-order phase transition temperature shows up in the strong frequency dependence of $\varepsilon'_r$ above this temperature. Below it, the dielectric relaxation decreases rapidly.

Similarly to samples with Schottky-type metal–semiconductor contacts, the huge values of $\varepsilon_r'$ at low frequencies can be attributed to the local formation of capacitive layers between the semiconducting sample and its metallic regions. This kind of response was originally treated by Maxwell[15] and Wagner[16] and is usually modeled by electromagnetic network analysis.[8] In our sample we have to account for both the internal (or intrinsic) metal-semiconductor interfaces, whose total contribution to the permittivity is magnetically controllable, and the contacts at the surface (variations in the nature of the electrodes cause a frequency shift of the $\varepsilon_r'$ vs. frequency curves, due to the expected Maxwell-Wagner effect of the electrodes). When adding the inductivity of the electrical connections or of the sample itself even dielectric resonance may be excited in this complex circuit. This is obvious from the spectra $\varepsilon_r'$ and $\sigma' = \varepsilon_o\omega\varepsilon_r''$ versus frequency, shown in Fig. 3 for $10^4 \leq f \leq 10^7$ Hz, which we obtained with a dielectric spectrometer Solartron 1260 with interface 1296 at 294 K. At low frequencies, $f < 6 \cdot 10^5$ Hz, we encounter a large Maxwell-Wagner-type dispersion step of the permittivity, $\Delta\varepsilon_r' \approx 1.3 \cdot 10^5$. It turns into a resonance curve centered at $f \approx 10^6$ Hz, and reveals a negative overshoot as is typical of resonance-like behavior, but has rarely been observed in dielectrics in the low-$f$ regime. It recovers to small positive values, $\varepsilon_r' \approx 200$, above $f \approx 10^7$ Hz. As expected from Kramers-Kronig relationships this feature comes together with a peak of the imaginary part of the permittivity $\varepsilon_r''$, or – more adequately – of the real part of the conductivity, $\sigma' = \varepsilon_o\omega\varepsilon_r''$ (solid symbols in Fig. 3). Such an anomalous resonance curve confirms this picture of interfacial polarization produced in the phase segregated regime.

On the other hand, the observed influence of the magnetic field on the dielectric response is in full agreement with such a phase segregated scenario. The application of a magnetic field has a strong effect above $T_C$, but very little below it, indicating that the magnetically dependent dielectric species exist only above this temperature.

Also, the reduction of the magnetic dependence of $\varepsilon'_r$ at higher temperatures is consistent with the weakening of the phase separation regime, that in $La_{2/3}Ca_{1/3}MnO_3$ is present only in a certain temperature window, $T_C < T < T^* \sim 350$ K.[16, 17]

From studies made on other CMR perovskites similar to $La_{2/3}Ca_{1/3}MnO_3$ it is well known that the application of a magnetic field above $T_C$ makes the magnetic regions grow, enhancing conductivity and giving rise to CMR.[19] At this point it is worth noting the qualitative similarities between the temperature dependences of resistivity and $\varepsilon'_r$, both with and without applied magnetic field (Fig. 4). This feature clearly points to a relationship between both magnitudes. In fact, for interfacial polarization, the measured $\varepsilon'_r$ depends on conductivity.[7] This behavior scales also with the temperature dependences of the intensity of neutron diffuse scattering at the peak positions of charge ordering modulation wave vectors[20] and of the polaron peak,[21] both measured in $La_{0.7}Ca_{0.3}MnO_3$, which reinforces the indication of a common physical origin. Given that $La_{2/3}Ca_{1/3}MnO_3$ is a CMR material, it seems that its magnetic field-dependent conductivity is the factor that underlies much of the observed magnetic field dependence of $\varepsilon'_r$.

In summary, we report a strong magnetic field dependence in the dielectric constant of $La_{2/3}Ca_{1/3}MnO_3$ (a well known non-ferroelectric CMR material) at room temperature and for low magnetic fields. Further work is to be done to explore the rich dielectric phenomenology arising from electronic phase separation in this and related systems.

We wish to acknowledge the financial support from DGICYT, Ministerio de Educación y Ciencia of Spain and UE (FEDER), under project MAT2004-05130.

**Figure captions**

Figure 1: Real part of the complex relative dielectric permittivity, $\varepsilon'_r$, of $La_{2/3}Ca_{1/3}MnO_3$ versus temperature at selected frequencies under no magnetic field. The temperature of the maximum of the different curves, $T_m$, does not change with frequency. The frequency dependence of $\varepsilon'_r$ is more pronounced for $T > T_m$. Data normalized to 235 K, 0 T and $1.5 \cdot 10^5$ Hz ($\varepsilon'_r = 7\,090$). The lines are guides to the eye.

Figure 2: Main frame: Temperature dependence of $\varepsilon'_r$ at $1.5 \cdot 10^7$ Hz under an applied magnetic field of 0.5 T, compared with the zero-field case. The lines are guides to the eye. Inset: magnetodielectric effect in $La_{2/3}Ca_{1/3}MnO_3$. The percentage variation at 0.5 T is more than 35% for temperatures above $T_C$.

Figure 3: ac permittivity $\varepsilon'_r$ and conductivity $\sigma'$ of $La_{2/3}Ca_{1/3}MnO_3$ measured at 294 K in zero external magnetic field at frequencies $10^4 \leq f \leq 10^7$ Hz. A Maxwell-Wagner-type resonance anomaly is observed around $f_0 \approx 10^6$ Hz.

Figure 4: dc resistivity of $La_{2/3}Ca_{1/3}MnO_3$ versus temperature (a) under no applied magnetic field and (b) under a magnetic field of 0.5 T. Data normalized to 235 K, 0 T ($\rho = 0.028\ \Omega$ cm). The temperature dependence of $\varepsilon'_r$ versus temperature (at $1.5 \times 10^5$ Hz)

under the same magnetic fields is included in each of these plots for comparison. The lines are guides to the eye.

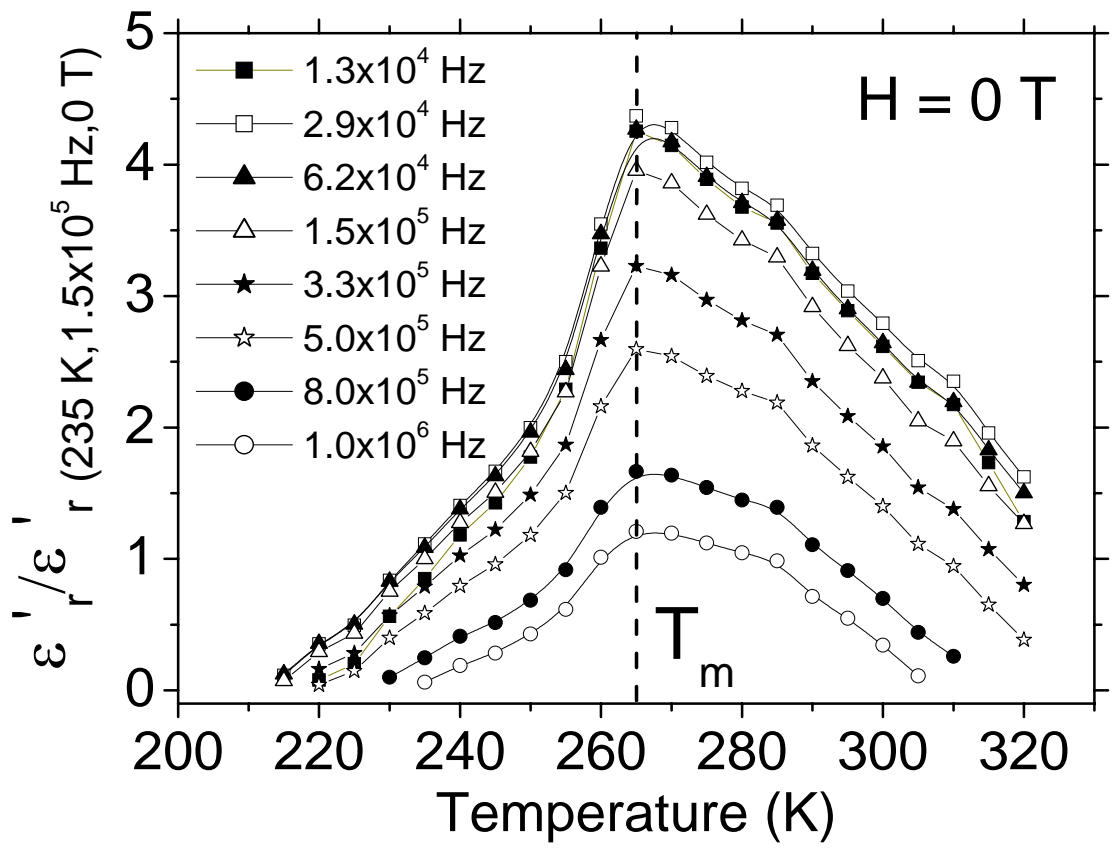

Figure 1

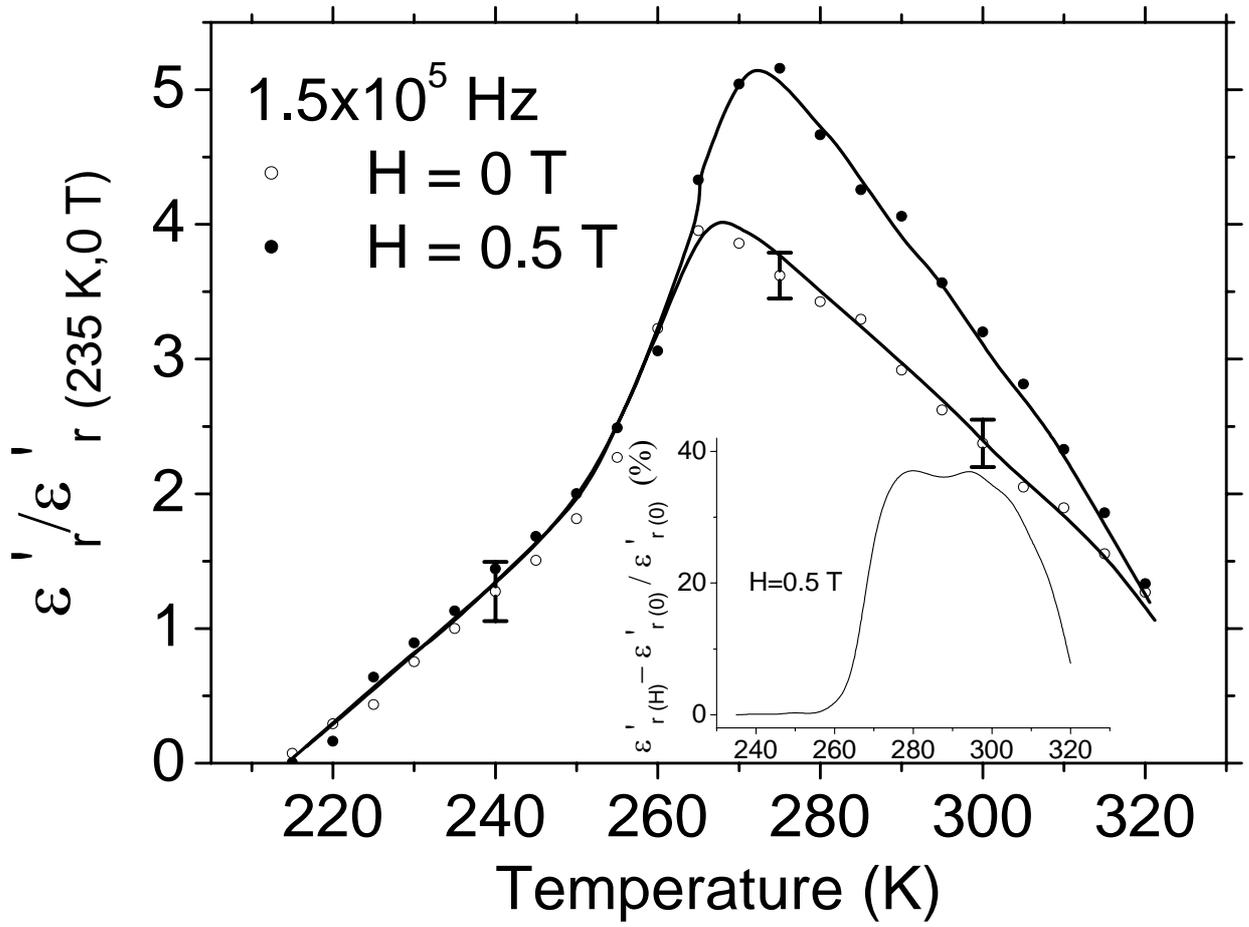

Figure 2

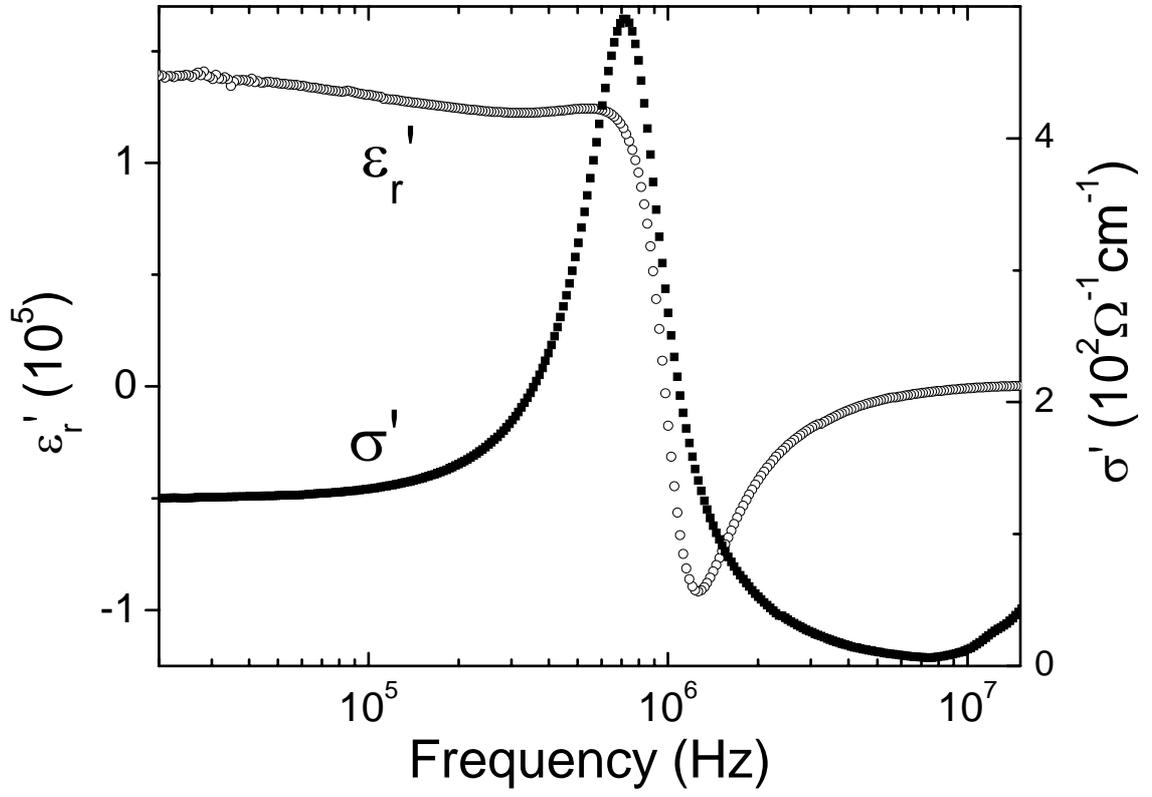

Figure 3

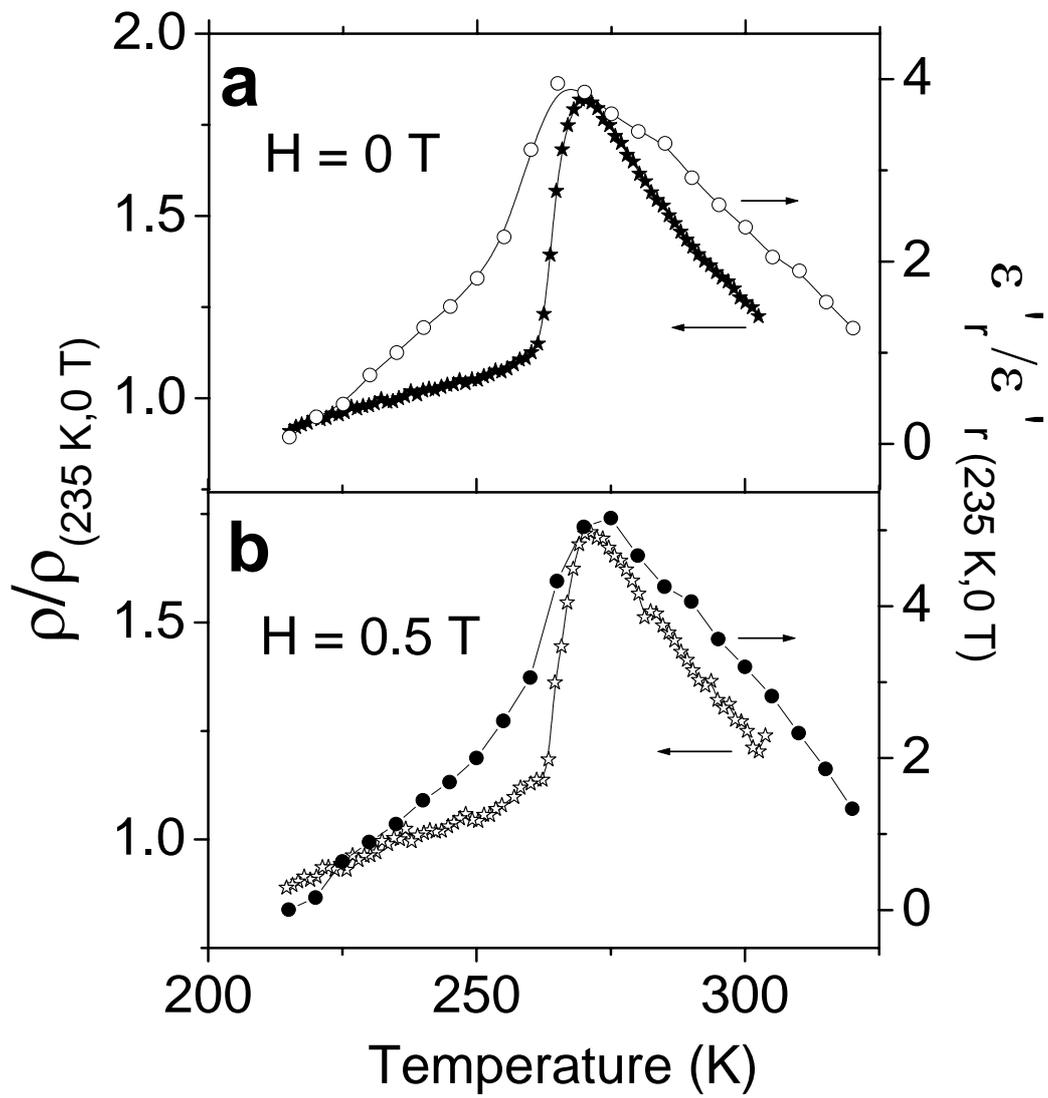

Figure 4